\documentclass[11pt]{article}
\usepackage[english]{babel}
\usepackage{amsmath,amssymb,mathtools,amsthm}
\allowdisplaybreaks

\newcommand{\1}{{\rm 1\hspace*{-0.4ex}%
\rule{0.1ex}{1.52ex}\hspace*{0.2ex}}}
 \expandafter\let\csname
equation*\endcsname\relax \expandafter\let\csname endequation*\endcsname\relax
\newcommand{\SE}{Schr\"{o}dinger equation}
\newcommand{\reduce}{\texttt{REDUCE}}

\newcommand{\rank}{\mathrm{rank}}
\newcommand{\half}{\frac{1}{2}}
\newcommand{\ud}{\mathrm{d}}

\newcommand{\R}{\mathbb{R}}

\renewcommand{\epsilon}{\varepsilon}
\renewcommand{\imath}{\mathrm{i}}
\newcommand{\abs}[1]{\left|#1\right|}
\def\dddot#1{\mathinner{\buildrel\vbox{\kern5pt\hbox{...}}\over{#1}}}
\newcommand{\di}{\displaystyle}
\newcounter{rmk}
\renewcommand{\thermk}{\arabic{rmk}}
\newenvironment{remark}%
{\refstepcounter{rmk}\noindent \textbf{Remark
\thermk:}}{\hfill$\blacksquare$\\\par \noindent}

\author{G.~Gubbiotti$^1$ \and M.~C.~Nucci$^2$}
\title{Quantization of the dynamics of a particle on a double cone
 by preserving Noether symmetries}
\date{$^1$Dipartimento di Matematica e Fisica, Universit\`{a} degli Studi Roma Tre,
 \& INFN Sezione di Roma Tre, 00146 Roma, Italy\\ [0.2cm]
$^2$Dipartimento di Matematica e Informatica, Universit\`a degli Studi di
Perugia, \& INFN Sezione di Perugia, 06123 Perugia, Italy}

\begin{document}

\maketitle

\begin{abstract}
The classical quantization of the motion of a free particle  and that of an
harmonic oscillator on a double cone are achieved by a quantization scheme [M.~C.
Nucci,  Theor. Math. Phys. 168 (2011) 994],  that preserves the
Noether point symmetries of the underlying Lagrangian in order to construct the
Schr\"odinger equation. The result is different from that given in [K.~Kowalski, J.~Rembiel\'{n}ski,  Ann. Phys. 329 (2013) 146].
A comparison of the different outcomes is provided.
\end{abstract}

\section{Introduction}

In \cite{c-iso} it was inferred that Lie symmetries should be preserved if a
consistent quantization is desired. In \cite{Goldstein80} [ex. 18, p. 433] an
alternative Hamiltonian for the simple harmonic oscillator was presented. It is
obtained by applying a nonlinear canonical transformation to the classical
Hamiltonian of the harmonic oscillator. That alternative Hamiltonian was used
in \cite{Nucci2013} to demonstrate what nonsense the usual quantization
schemes\footnote{Such as normal-ordering \cite{Bjorken1964,Louisell1990} and
Weyl quantization \cite{Weyl1927}.} produce. In \cite{gallipoli10} a
quantization scheme that preserves the Noether symmetries was proposed and
applied  to Goldstein's example in order to derive the correct Schr\"odinger
equation. In \cite{Bregenz11}  the same quantization scheme was applied in
order to quantize the second-order Riccati equation, while in \cite{Nucci2012}
the quantization of the dynamics of a charged particle in a uniform magnetic
field in the plane and Calogero's goldfish system were achieved. In
\cite{AGMP09} the same method yielded the Schr\"odinger equation of  an
equation related to a
 Calogero's goldfish, and in \cite{PMNP13} that of two nonlinear
equations somewhat related to the Riemann problem \cite{Sierra12}.
 In \cite{GN_liensch}, and \cite{GN_LienII} it was shown that the preservation
 of the Noether symmetries
straightforwardly yields the Schr\"odinger equation  of a Li\'{e}nard I
nonlinear oscillator in the momentum space  \cite{Senth2012},
and that of a family of Li\'{e}nard II nonlinear oscillators \cite{Partha2013},
 respectively.\\

 If a system of second-order equations is considered, i.e.
\begin{equation}
{\ddot{\mathbf{x}}}(t)= \mathbf{F}(t,\mathbf{x},\mathbf{\dot{x}}), \quad
\mathbf{{x}}\in\R^N, \label{systsec}
\end{equation}
that comes from a variational principle with a Lagrangian of first order, then
the method that was first proposed in \cite{gallipoli10} consists of the
following steps:
\begin{description}
\item[Step I.] Find the Lie symmetries of the Lagrange equations
$$\Upsilon=W(t,\mathbf{x})\partial_t+\sum_{k=1}^{N}W_k(t,\mathbf{x})\partial_{x_k}$$
\item[Step II.]  Among them find the Noether  symmetries
$$\Gamma=V(t,\mathbf{x})\partial_t+\sum_{k=1}^{N}V_k(t,\mathbf{x})\partial_{x_k}$$
This may require searching for the Lagrangian yielding the maximum possible
number of Noether  symmetries
\cite{laggal,CP07Rao1JMP,nuctam_1lag,nuctam_3lag}.
\item[Step III.]  Construct the Schr\"odinger equation\footnote{We assume
 $\hbar=1$ without loss of generality.} admitting these Noether
symmetries as Lie symmetries, namely
\begin{equation}2\imath\Psi_t+\sum_{k,j=1}^{N} f_{kj}(\mathbf{x})\Psi_{x_jx_k}+
\sum_{k=1}^{N}h_k(\mathbf{x})\Psi_{x_k}+f_0(\mathbf{x})\Psi=0
\label{sch}\end{equation} with Lie symmetries
$$\Omega=V(t,\mathbf{x})\partial_t+\sum_{k=1}^{N}V_k(t,\mathbf{x})\partial_{x_k}
+G(t,\mathbf{x},\Psi)\partial_{\Psi}$$ without adding any other symmetries
apart from the two symmetries that are present in any linear homogeneous
partial differential equation\footnote{In the following we will refer to those
two symmetries as the homogeneity and linearity symmetries.}, namely
$${\Psi}\partial_{\Psi}, \quad \quad \alpha(t,\mathbf{x})\partial_{\Psi},$$
where $\alpha=\alpha(t,\mathbf{x})$ is any solution of the Schr\"odinger
equation \eqref{sch}.
\end{description}
\strut\hfill\\ If the system \eqref{systsec}  is linearizable by a point
transformation, and it possesses the maximal number of admissible Lie point
symmetries, namely $N^2+4N+3$, then in \cite{Gonzalez1983,Gonzalez1988} it was
proven that the maximal-dimension Lie symmetry algebra of a system of $N$
equations of second order is isomorphic to $sl(N+2,\R)$, and that the
corresponding Noether symmetries generate a $(N^2+3N+6)/2$-dimensional Lie
algebra $g^V$ whose structure (Levi-Mal\'cev decomposition and realization by
means of a matrix algebra) was determined. It was also proven that the
corresponding linear system is
\begin{equation}
\mathbf{y''}(s)+2A_1(s)\cdot\mathbf{y}'(s)+
A_0(s)\cdot\mathbf{y}(s)+\mathbf{b}(s)=0, \label{linsys}
\end{equation}
 with the condition
\begin{equation}
A_0(s)=A_1'(s)+A_1(s)^2+a(s)\1\,,
\end{equation}
where $A_0,A_1$ are $N\times N$ matrices, and $a$ is a scalar function.\\
\\Consequently if  system \eqref{systsec} admits $sl(N+2,\R)$ as Lie symmetry algebra
then in \cite{GN_liensch} we reformulated the algorithm that yields the
Schr\"odinger equation as follows:
\begin{description}
\item[Step 1.] Find the linearizing transformation which
does not change the time, as prescribed in non-relativistic quantum mechanics.
\item[Step 2.]
Derive the Lagrangian  by applying the linearizing transformation to the
standard Lagrangian of the corresponding linear system \eqref{linsys}, namely
the one that admits the maximum number of Noether symmetries\footnote{In
\cite{Gonzalez1988}  it was shown that any diffeomorphism between two systems
of second-order differential equations takes Noether symmetries into Noether
symmetries, and therefore the Lagrangian is unique up to a diffeomorphism.}.
\item[Step 3.] Apply the linearizing transformation
 to the Schr\"{o}dinger equation  of the
corresponding classical linear problem. This yields the Schr\"odinger~ equation
corresponding to system \eqref{systsec}.
\end{description}
This quantization is consistent with the classical properties of the system,
namely  the Lie symmetries of the obtained Schr\"{o}dinger equation correspond
to the Noether symmetries admitted by the Lagrangian of system
\eqref{systsec}.\\
\strut\hfill\\
In  \cite{Kowalski2013} the problem of the quantization of the dynamics of a
particle constrained on a double cone was considered. \\A double cone of
opening angle $2 \alpha$ with $\alpha\in (0,\pi/2)$ is given by the cartesian
equation:
\begin{equation}
x^2 + y^2 -\cot^2(\alpha) z^2=0.
\end{equation}
A regular parametrization (except at the vertex $(0,0,0)$)
for this surface is given by:
\begin{subequations}
\begin{align}
x &= r \sin (\alpha) \cos(\phi),\\
y &= r \sin (\alpha) \sin (\phi),\\
z &= r \cos (\alpha)
\end{align}
\label{conepar}
\end{subequations}
with $r\in \R$ and $\phi \in \mathopen{[}0,2\pi\mathclose{)}$.\\
Consequently a particle of mass $m$ constrained on a cone has the natural
Lagrangian:
\begin{equation}
L =  \half m\left(\dot{r}^{2} +  \sin^2 (\alpha) r^2 \dot{\phi}^2\right)- V(r,\phi),
\label{lagrcone}
\end{equation}
where $V(r,\phi)$ is the potential energy.

\begin{remark}  Without loss of generality we assume $m=1$.
Also, since $\sin(\alpha)$ is a constant, we introduce another constant $k\in
(0,1)$ such that $\sin(\alpha)=k$ in order to have neater expressions.
\end{remark}

In \cite{Kowalski2013} the authors addressed the quantization of two particular
cases of the Lagrangian \eqref{lagrcone}, namely the free particle, with
Lagrangian\footnote{$V(r,\phi)=0$.}:
\begin{equation}
L_{\text{f}} = \half \left(\dot{r}^{2} +  k^2 r^2 \dot{\phi}^2\right),
\label{lagrconefree}
\end{equation}
and the radial harmonic oscillator, with Lagrangian\footnote{$V(r,\phi)=
\frac{1}{2}\omega^2 r^2$.}:
\begin{equation}
L_{\text{ho}} = \half \left(\dot{r}^{2} +  k^2 r^2 \dot{\phi}^2\right) - \half\omega^2 r^2.
\label{lagrconeho}
\end{equation}
The radial harmonic oscillator was also considered in \cite{Briha} as the classical motion of a particle on a
cone under the influence of a central potential.

In this paper we apply the quantization algorithm that preserves the Noether
symmetries to those two cases, and also determine the eigenvalues and the
eigenfunctions of the obtained Schr\"odinger equations. We compare the results of our
quantization method with those obtained in \cite{Kowalski2013}, and explain the
differences that are due to symmetry breaking.

\section{Quantization of a free particle on the cone}
The Lagrangian  equations corresponding to the Lagrangian \eqref{lagrconefree}
are:
\begin{equation} \left\{
\begin{array}{rcl}
\ddot{r} &=&k^2 r \dot{\phi}^2,\\[0.3cm]
\ddot{\phi} &=&- \di 2\frac{\dot{r} \dot{\phi}}{r}.
\end{array}  \right. \label{eqconefree}
\end{equation}
Using the \reduce~programs \cite{Nucci1996} we find that this system  admits a
fifteen-dimensional Lie point symmetry algebra, isomorphic to
$\mathrm{sl}(4,\R)$,  generated by the following operators:
\begin{eqnarray}
\Gamma_1 &=&\cos(k \phi) r ( t \partial_{t}  + r \partial_{r}),
\nonumber \\
\Gamma_2 &=&\cos(k \phi) r \partial_{t},
\nonumber \\
\Gamma_3 &=&\sin(k \phi) r ( t \partial_{t}  + r \partial_{r}),
\nonumber \\
\Gamma_4 &=&\sin(k \phi) r \partial_{t},
\nonumber \\
\Gamma_5 &=&t ( t \partial_{t}  + r \partial_{r}),
\nonumber \\
\Gamma_6 &=&\half r \partial_{r} + t\partial_{t},
\nonumber \\
\Gamma_7 &=&\partial_{t},
\nonumber \\
\Gamma_8 &=&t \left(\cos(k \phi)  \partial_{r} - \frac{1}{k r}\sin(k \phi)
\partial_{\phi}\right), \label{symmconefree}
\\
\Gamma_9 &=&\cos(k \phi)  \partial_{r} -\frac{1}{k r} \sin(k \phi)
\partial_{\phi},
\nonumber \\
\Gamma_{10} &=&t \left(\sin(k \phi)  \partial_{r} + \frac{1}{k r}\cos(k \phi)
\partial_{\phi}\right),
\nonumber \\
\Gamma_{11} &=&\sin(k \phi)  \partial_{r} +\frac{1}{k r} \cos(k \phi)
\partial_{\phi},
\nonumber \\
\Gamma_{12} &=& r\partial_{r},
\nonumber \\
\Gamma_{13} &=& k r \cos(2 k \phi)  \partial_{r} - \sin(2 k \phi)
\partial_{\phi},
\nonumber \\
\Gamma_{14} &=& k r \sin(2 k \phi)  \partial_{r} + \cos(2 k \phi)
\partial_{\phi},
\nonumber \\
\Gamma_{15} &=&\partial_{\phi}.\nonumber
\end{eqnarray}
Consequently system  \eqref{eqconefree} is linearizable
\cite{Gonzalez1983,Gonzalez1988}, and in order to quantize it we follow
the three Steps 1,2,3 \cite{GN_liensch} as recalled in the Introduction.\\

\noindent {\bf Step 1.} We have to find a linearizing transformation which does
not alter the time $t$. In \cite{Soh2001} it
was determined that the linearizing transformation can be found by means of a
four-dimensional subalgebra of type $A_{4,5}^{1,1} = \langle
X_{1},X_{2},X_{3},X_{4}\rangle$ in the Mubarakzyanov classification
\cite{Mubarakzyanov1963a,Mubarakzyanov1966}
with commutation relations:
\begin{equation}
[X_{i},X_{j}] = 0, \quad [X_{i},X_{4}] = X_{i}, \quad (i,j=1,2,3) \label{a4511}
\end{equation}
such that $1\leq\rank[X_{1},X_{2},X_{3},X_{4}]\leq 3$, and whose canonical form
is:
\begin{equation}
\begin{gathered}
X_{1} = \partial_{\tau}, \quad X_{2} = \partial_{u}, \quad X_{3} =
\partial_{v},
\\
X_{4} = \tau \partial_{\tau} + u \partial_{u}
+v \partial_{v}.
\label{cana4511}
\end{gathered}
\end{equation}
We found that the operators $\Gamma_{7}$, $\Gamma_{9}$, $\Gamma_{11},
\Gamma_{16}$, with $\Gamma_{16} = \Gamma_{6} + \half \Gamma_{12}$, generate
such a subalgebra. Consequently the following transformation:
\begin{equation}
\tau = t, \quad u = r \cos(k\phi), \quad v = r\sin(k\phi).
\label{transffree}
\end{equation}
takes system \eqref{eqconefree} into the following linear
system\footnote{Namely the Lagrangian equations of a two-dimensional free
particle.}:
\begin{equation}
\ddot{u} = 0, \quad \ddot{v} = 0.
\label{linfree}
\end{equation}
Then the general solution of system \eqref{eqconefree} is:
\begin{eqnarray}
r &=& \pm \sqrt{(c_{1}t+c_{2})^2 + (c_{3}t + c_{4})^2}, \nonumber
\\
\tan\left(k\phi\right) &=& \frac{c_{1}t + c_{2}}{c_{3}t + c_{4}},
\label{eqconefreesol}
\end{eqnarray}
with $c_{i} \,(i=1,4)$ arbitrary constants.\\

\noindent {\bf Step 2.} The Lagrangian \eqref{lagrconefree}  admits eight
Noether symmetries, i.e.: $\Gamma_{5}$, $\Gamma_{6}$, $\Gamma_{7}$, $\Gamma_{8}$,
$\Gamma_{9}$, $\Gamma_{10}$,
$\Gamma_{11}$, $\Gamma_{15}$. \\

\noindent {\bf Step 3.} The Schr\"{o}dinger equation of the two-dimensional
free particle in the variables $(u,v)$ is
\begin{equation}
2 \imath \psi_{t} + \psi_{uu} + \psi_{vv} = 0,
\end{equation}
with $\psi=\psi(t,u,v)$. If we  apply the transformation \eqref{transffree},
then we obtain that the  \SE~  corresponding to system \eqref{eqconefree}
is\footnote{Introducing $\hbar$ into \eqref{schrconefree}, i.e.
$$  2 \imath\hbar \psi_{t} + \hbar^2\left(\psi_{rr}
+ \frac{1}{r}\psi_{r} + \frac{1}{k^2 r^2} \psi_{\phi\phi}\right)=0,$$ and
performing the classical limit \cite{Landau} yields the
Hamilton-Jacobi equation for system \eqref{eqconefree}.}
\begin{equation}
2 \imath \psi_{t} + \psi_{rr} + \frac{1}{r}\psi_{r} + \frac{1}{k^2 r^2}
\psi_{\phi\phi}=0. \label{schrconefree}
\end{equation}
 Using the \reduce~programs \cite{Nucci1996} we find that
its Lie point symmetries are generated by the following operators:
\begin{eqnarray}
\Lambda_{1} &=& \Gamma_{15},\nonumber \\
\Lambda_{2} &=& \Gamma_{8} + \imath k^2 r \cos(k\phi)\psi\partial_{\psi},
\nonumber \\
\Lambda_{3} &=& \Gamma_{9},
\nonumber\\
\Lambda_{4} &=& \Gamma_{10} + \imath k^2 r \sin(k\phi)\psi\partial_{\psi},
\\
\Lambda_{5} &=& \Gamma_{11},
\nonumber\\
\Lambda_{6} &=& \Gamma_{5} +\half \left(\imath r^2 -2
t\right)\psi\partial_{\psi},
\nonumber\\
\Lambda_{7} &=& \Gamma_{6},
\nonumber\\
\Lambda_{8} &=& \Gamma_{7},
\nonumber\\
\Lambda_{9} &=& \psi\partial_{\psi}, \quad \Lambda_{\Psi} =
\Psi\partial_{\psi},\nonumber
\end{eqnarray}
where $\Psi(t,r,\phi)$ is any solution of equation \eqref{schrconefree}.

\medskip

In order to determine the  radial Schr\"odinger equation, we consider the two-dimensional abelian Lie subalgebra generated by the operators
\begin{eqnarray}\Lambda_{\epsilon} &=& \Lambda_{8} - \imath\epsilon\Lambda_{9}=\partial_t-\imath\epsilon\psi\partial_{\psi},\nonumber\\
\Lambda_{\beta} &=& \Lambda_{1} - \imath p \Lambda_{9}=\partial_{\phi}-\imath p\psi\partial_{\psi},\end{eqnarray}
with $\epsilon$ and $p$ are  arbitrary constants.
 Then solving the corresponding invariant surface condition \cite{Olver1986} yields the following invariant solution
\begin{equation}
\psi(t,r,\phi)=R(r) e^{-\imath (\epsilon t + p \phi)},
\label{invsolfree}
\end{equation}
that replaced into \eqref{schrconefree}  gives rise to the following radial Schr\"odinger equation:
\begin{equation}
R'' + \frac{1}{r}R'
+\left(2\epsilon-\frac{p^2}{k^2r^2}\right)R=0,\label{radfree}
\end{equation}
where prime denotes derivative by $r$.\\
After imposing  $p$ to be an integer since the invariant solution \eqref{invsolfree} must be periodic of period
$2\pi$, we obtain that the only bounded solution of equation \eqref{schrconefree}  is given in terms of the Bessel function of the first kind $J_{\mu}$, i.e.:
\begin{equation}
\psi_{p,\epsilon} =
J_{\frac{|p|}{k}}\left(\sqrt{2\epsilon}\, |r|\right)
e^{-\imath (\epsilon t + p \phi)},
\end{equation}
with the additional condition $\epsilon>0$.

\section{Radial harmonic oscillator on the cone}

The Lagrangian equations corresponding to the Lagrangian \eqref{lagrconeho} are:
\begin{equation} \left\{
\begin{array}{rcl}
\ddot{r} &=&k^2 r \dot{\phi}^2 - \omega^2 r,\\
\ddot{\phi} &=&- \di 2\frac{\dot{r} \dot{\phi}}{r}.
\end{array}  \right. \label{eqhocone}
\end{equation}
Using the \reduce~programs \cite{Nucci1996} we find that this system  admits a
fifteen-dimensional Lie point symmetry algebra, isomorphic to
$\mathrm{sl}(4,\R)$,  generated by the following operators:
\begin{eqnarray}
\Xi_{1} &=& \cos(k \phi) r \left( \cos( \omega t)  \partial_{t} - \omega r
\sin( \omega t)  \partial_{r}\right),
\nonumber\\
\Xi_{2} &=& \cos(k \phi) r \left(\sin( \omega t)  \partial_{t} + \omega r \cos(
\omega t)  \partial_{r}\right),
\nonumber\\
\Xi_{3} &=& \sin(k \phi) r \left( \cos( \omega t)  \partial_{t} -  \omega r
\sin( \omega t)  \partial_{r} \right),
\nonumber\\
\Xi_{4} &=& \sin(k \phi) r \left( \sin( \omega t)  \partial_{t} + \omega r
\cos( \omega t)  \partial_{r}\right),
\nonumber\\
\Xi_{5} &=& \cos(2 k \phi) r \partial_{r}  -  \frac{1}{k}\sin(2 k \phi)
\partial_{\phi},
\nonumber\\
\Xi_{6} &=& \sin(2 k \phi) r \partial_{r} + \frac{1}{k} \cos(2 k \phi)
\partial_{\phi},
\nonumber\\
\Xi_{7} &=& r \partial_{r},
\nonumber\\
\Xi_{8} &=& \partial_{\phi},\label{symmho}
\\
\Xi_{9} &=& \partial_{t},
\nonumber\\
\Xi_{10} &=& \cos(2  \omega t)  \partial_{t} -  \omega \sin(2  \omega t) r
\partial_{r},
\nonumber\\
\Xi_{11} &=& \sin(2  \omega t)  \partial_{t} +  \omega \cos(2  \omega t) r
\partial_{r},
\nonumber\\
\Xi_{12} &=& \cos( \omega t) \left( \cos(k \phi)  \partial_{r} - \frac{1}{k r}
\sin(k \phi)  \partial_{\phi}\right),
\nonumber\\
\Xi_{13} &=& \sin( \omega t) \left( \cos(k \phi)  \partial_{r} -  \frac{1}{k r}
\sin(k \phi)  \partial_{\phi}\right),
\nonumber\\
\Xi_{14} &=& \cos( \omega t) \left(\sin(k \phi)  \partial_{r} + \frac{1}{k r}
\cos(k \phi)  \partial_{\phi}\right),
\nonumber\\
\Xi_{15} &=& \sin( \omega t) \left(\sin(k \phi)  \partial_{r} + \frac{1}{k r}
\cos(k \phi)  \partial_{\phi}\right). \nonumber
\end{eqnarray}
Consequently system  \eqref{eqhocone} is linearizable
\cite{Gonzalez1983,Gonzalez1988}, and in order to quantize it we follow
the three Steps 1,2,3 \cite{GN_liensch} as recalled in the Introduction.\\

\noindent {\bf Step 1.}   The transformation
\eqref{transffree} applied to \eqref{eqhocone} yields:
\begin{equation}
\ddot{u} + \omega^2 u = 0,\quad \ddot{v}  + \omega^2 v= 0,
\end{equation}
namely the equations of a two-dimensional linear harmonic oscillator. Therefore, the
general solution of \eqref{eqhocone} is: \begin{eqnarray}\!\!\!\!\!\!\!\! r \!\!\!\!&=& \!\!\!\!\pm
\sqrt{\Big(c_{1}\cos\left(\omega t\right) +c_{2}\sin\left(\omega
t\right)\Big)^2 + \Big(c_{3}\cos\left(\omega t\right) +c_{4}\sin\left(\omega
t\right)\Big)^2}, \nonumber
\\
\!\!\!\!\tan(k\phi) \!\!\!\!&=& \!\!\!\!\frac{c_{1}\cos\left(\omega t\right) +c_{2}\sin\left(\omega
t\right)}{c_{3}\cos\left(\omega t\right) +c_{4}\sin\left(\omega
t\right)}. \label{eqconehosol}
\end{eqnarray}

\noindent {\bf Step 2.} The Lagrangian \eqref{lagrconeho} admits eight Noether
symmetries, namely $\Xi_{i}$ with $i=8,\ldots,15$ in \eqref{symmho}.\\

\noindent {\bf Step 3.} The Schr\"{o}dinger equation for a two-dimensional
linear harmonic oscillator in the variables $(u,v)$  and wave function $\psi$ is:
\begin{equation}
2 \imath \psi_{t} + \psi_{uu} + \psi_{vv} -\omega^2 (u^2 + v^2)\psi= 0.
\end{equation}
If we apply the transformation \eqref{transffree}, then the Schr\"{o}dinger
equation corresponding to system \eqref{eqhocone} is\footnote{Introducing $\hbar$
into \eqref{schrhocone}, i.e.
$$ 2 \imath\hbar \psi_{t} + \hbar^2\left(\psi_{rr} + \frac{1}{r}\psi_{r} +
\frac{1}{k^2 r^2} \psi_{\phi\phi}\right) - \omega^2 r^2\psi=0,$$ and
performing the classical limit \cite{Landau} yields the
Hamilton-Jacobi equation for  system \eqref{eqhocone}.}:
\begin{equation}
2 \imath \psi_{t} + \psi_{rr} + \frac{1}{r}\psi_{r} +
\frac{1}{k^2 r^2} \psi_{\phi\phi} - \omega^2 r^2\psi=0.
\label{schrhocone}
\end{equation}
 Using the \reduce~programs \cite{Nucci1996} we find that
its Lie point symmetries are generated by the following operators:
\begin{eqnarray}
\Omega_{1} &=& \Xi_{8},
\nonumber\\
\Omega_{2} &=& \Xi_{10} +
\omega  \left(\sin(2 \omega t) - 2\imath  \cos(2 \omega t) \omega r^2\right)\psi \partial_{\psi},
\nonumber\\
\Omega_{3} &=& \Xi_{11} -
\omega  \left(\cos(2 \omega t) + 2\imath  \sin(2 \omega t) \omega r^2\right)\psi\partial_{\psi},
\nonumber\\
\Omega_{4} &=& \Xi_{9},
\nonumber\\
\Omega_{5} &=& \Xi_{14} - \imath \omega r \sin(\omega t) \sin(k\phi) \psi
\partial_{\psi},
\\
\Omega_{6} &=& \Xi_{15} + \imath \omega r \cos(\omega t) \cos(k\phi) \psi
\partial_{\psi},
\nonumber\\
\nonumber\Omega_{7} &=& \Xi_{12} - \imath \omega r \sin(\omega t) \cos(k\phi)
\psi
\partial_{\psi},
\nonumber\\
\Omega_{8} &=& \Xi_{13} + \imath \omega r \cos(\omega t) \cos(k\phi) \psi
\partial_{\psi},
\nonumber\\
\Omega_{9} &=& \psi \partial_{\psi}, \quad \Omega_{\Psi} = \Psi
\partial_{\psi},\nonumber
\end{eqnarray}
where ${\Psi}=\Psi(t,r,\phi)$ is any solution of \eqref{schrhocone}.

\medskip

Equation \eqref{schrhocone} is the same Schr\"odinger equation that was obtained  in \cite{Furtado,Hashi,Mako} in the case that $r \in (0,\infty)$. Therefore the eigenfunctions of
equation \eqref{schrhocone} are given in terms of the associated Laguerre polynomials $L_n^{\mu}$, i.e.:
\begin{equation}
\psi_{n,p}=e^{-\imath E_n t+\imath p\phi}|r|^{|p|/k}e^{-(1/2)\omega r^2}L_n^{|p|/k}(\omega r^2), \quad \quad (n \in \mathbb{N}, p \in \mathbb{Z})
\end{equation}
with eigenvalues
\begin{equation}
E_n=\omega\left(2n+\frac{|p|}{k}+1\right).\label{eigen}
\end{equation}

\section{Comparison of the different outcomes}

The quantization of the harmonic oscillator on a cone with one nappe has been
studied in several papers, e.g. \cite{Furtado,Hashi,Mako}.

Our Schr\"odinger equation \eqref{schrhocone} is exactly the same that all the above authors derived by means of the Laplace-Beltrami operator, namely
 Noether symmetries of the classical Lagrangian systems are preserved as we shown.

In \cite{Kowalski2013}, the authors derived a different Schr\"odinger equation for both the free particle and the harmonic oscillator on the double cone.
Instead of using the Laplace-Beltrami operator they look for self-adjoint operators of the type
\begin{equation}
\hat{p}_{r} = -\imath \left(\partial_{r} +F(r)\right),
\end{equation}
with respect to the scalar product
\begin{equation}
\langle f, g\rangle =
\int_{0}^{2\pi}\!\!\!\int_{-\infty}^{\infty} f^{*}(r,\phi)g(r,\phi)
\abs{r} \ud r \ud \phi,
\label{l2cone}
\end{equation}
on the space of square integrable functions $f(r,\phi), g(r,\phi)$
on the cone. This yields that the self-adjoint operator is:
\begin{equation}
\hat{p}_{r} = -\imath \left(\partial_{r} +\frac{1}{2 r}\right),
\end{equation}
and consequently the  \SE~for the free particle is derived to be:
\begin{equation}
2\imath \psi_{t} + \psi_{rr}+\frac{1}{r}\psi_{r}
+\frac{1}{4k^2 r^2}\psi_{\phi\phi} -\frac{1}{4r^2}\psi = 0,
\label{schrkowfree}
\end{equation}
while the \SE~ for the harmonic oscillator becomes:
\begin{equation}
2\imath \psi_{t} + \psi_{rr}+\frac{1}{r}\psi_{r} +\frac{1}{4k^2
r^2}\psi_{\phi\phi} - \left(\frac{1}{4r^2} + \omega^2 r^2\right)\psi = 0.
\label{schrkowho}
\end{equation}
Both equations \eqref{schrkowfree} and \eqref{schrkowho} do not preserve the Noether symmetries of the free particle and the harmonic oscillator on the cone, respectively.
Indeed the Lie symmetries of equation \eqref{schrkowfree} are:
\begin{eqnarray}
\Upsilon_{1} &=& \Gamma_{15},
\\
\Upsilon_{2} &=& \Gamma_5 +\half \left(\imath r^2 -2
t\right)\psi\partial_{\psi},
\\
\Upsilon_3 &=& \Gamma_{6},
\\
\Upsilon_4 &=& \Gamma_7,
\\
\Upsilon_{5} &=& \psi \partial_{\psi}, \quad \Upsilon_{\Psi} =
\Psi\partial_{\psi},
\end{eqnarray}
with ${\Psi}=\Psi(t,r,\phi)$  any solution of \eqref{schrkowfree}, while those of equation \eqref{schrkowho} are:
\begin{eqnarray}
\Pi_{1} &=& \Xi_{8},
\\
\Pi_{2} &=& \Xi_{11}
- 
\omega (\cos(2 \omega t) + 2\imath \sin(2 \omega t) \omega r^2)\psi \partial_{\psi},
\\
\Pi_{3} &=& \Xi_{10}
+ 
\omega  (\sin(2 \omega t) - 2\imath  \cos(2 \omega t) \omega r^2)\psi\partial_{\psi},
\\
\Pi_{4} &=& \Xi_{9},
\\
\Pi_{5} &=& \psi \partial_{\psi}, \quad \Pi_{\Psi} =
\Psi\partial_{\psi},
\end{eqnarray}
with ${\Psi}=\Psi(t,r,\phi)$  any solution of \eqref{schrkowho}.
 Indeed,  the additional term $-\psi/4r^2$ in both equations \eqref{schrkowfree} and \eqref{schrkowho} breaks the symmetries, i.e. four out of eight symmetries are not preserved.
 Therefore, our eigenfunctions  are different from those obtained in \cite{Kowalski2013} although the boundary conditions are the same.
 Moreover, in the case of the harmonic oscillator the eigenvalues that were derived in \cite{Kowalski2013} are:
 \begin{equation}
E_n=\omega\left(2n+\frac{1}{2}\sqrt{1+\frac{4p^2}{k^2}}+1\right),
\end{equation}
 instead of those that we derived, i.e. \eqref{eigen}.

 Finally, we would like to underline that the eigenvalues \eqref{eigen} that we have obtained coincide with those derived by other authors, e.g. \cite{Furtado,Hashi,Mako}, if one nappe only is considered.

\section{Conclusions}

In this paper we have derived the Schr\"odinger equation for both the free particle and the harmonic oscillator on a double cone by requiring the preservation of the Noether symmetries of the classical problem.  Indeed the Noether symmetries admitted by the Lagrangian \eqref{lagrconefree} are the Lie symmetries of the Schr\"odinger equation \eqref{schrconefree}, and the Noether symmetries admitted by the Lagrangian \eqref{lagrconeho} are the Lie symmetries of the Schr\"odinger equation \eqref{schrhocone}. In particular the latter  coincides with the Schr\"odinger equation derived by other authors,  e.g. \cite{Furtado,Hashi,Mako} in the case of a single cone.  On the other hand, the Schr\"odinger equations for the same problems that were obtained in  \cite{Kowalski2013} do not preserve the Noether symmetries of the classical problem and therefore yield results quite different from the ones we have derived here. Further insight is needed especially from the experimentalists as stated in \cite{Carboncone}.

\section*{Acknowledgements}
\noindent
GG is supported by INFN IS-CSN4 {\em Mathematical Methods of Nonlinear Physics}.\\
MCN acknowledges the support of the Italian Ministry of University and
Scientific Research through PRIN 2010-2011, Prot. 2010JJ4KPA\_004, {\em Geometric
and analytic theory of Hamiltonian systems in finite and infinite dimensions}.

\end{document}